\documentclass[prl,aps,reprint,amsmath,amssymb,longbibliography, superscriptaddress]{revtex4-2}

\PassOptionsToPackage{bbgreekl}{mathbbol}

\usepackage{ajtex-revtex}
\usepackage{multirow}

\makeatletter 
    
\renewcommand\onecolumngrid{
\do@columngrid{one}{\@ne}%
\def\set@footnotewidth{\onecolumngrid}
\def\footnoterule{\kern-6pt\hrule width 1.5in\kern6pt}%
}

\renewcommand\twocolumngrid{
\def\footnoterule{
\dimen@\skip\footins\divide\dimen@\thr@@
\kern-\dimen@\hrule width.5in\kern\dimen@}
\do@columngrid{mlt}{\tw@}
}%

\makeatother

\renewcommand{\paragraph}[1]{\vspace{1em}\noindent\textbf{#1}}

\definecolor{ogreen}{RGB}{71,191,145}

\newcommand{\vct}[1]{\mathbf{#1}}

\newcommand{\quotes}[1]{``#1''}

\begin{document} 

\title{Finite-volume scheme for first-order viscoresistive relativistic magnetohydrodynamics}

\author{Ruben Lier}\email{rlier@uva.nl}

\affiliation{Institute for Theoretical Physics, University of Amsterdam, 1090 GL Amsterdam, The Netherlands}
\affiliation{Dutch Institute for Emergent Phenomena, 1090 GL Amsterdam, The Netherlands}
\affiliation{Institute for Advanced Study, University of Amsterdam, Oude Turfmarkt 147, 1012 GC Amsterdam, The Netherlands}

\author{Jay Armas}\email{j.armas@uva.nl}

\affiliation{Institute for Theoretical Physics, University of Amsterdam, 1090 GL Amsterdam, The Netherlands}
\affiliation{Dutch Institute for Emergent Phenomena, 1090 GL Amsterdam, The Netherlands}
\affiliation{Institute for Advanced Study, University of Amsterdam, Oude Turfmarkt 147, 1012 GC Amsterdam, The Netherlands}
\affiliation{Niels Bohr International Academy, The Niels Bohr Institute, University of Copenhagen,
Blegdamsvej 17, DK-2100 Copenhagen \O{}, Denmark}

\author{Oliver Porth}\email{o.j.g.porth@uva.nl}
\affiliation{Anton Pannekoek Institute, Science Park 904, 1098 XH, Amsterdam, The Netherlands}

\date{\today}

\begin{abstract}
We present a numerical implementation of dissipative relativistic magnetohydrodynamics based on Bemfica–Disconzi–Noronha–Kovtun (BDNK) theory, in which first-order corrections render the equations causal without introducing additional dynamical variables. We show how these corrections can be incorporated in a finite-volume scheme describing the coupled dissipation of energy, momentum, and magnetic field, with the latter treated as a one-form charge. While a minimal set of BDNK terms can convert diffusive equations into telegrapher-type equations, we find that in the ultra-relativistic limit an additional correction is required for the system to behave in a stable and causal manner. With this set of equations, we develop an efficient method for primitive-variable recovery and validate the implementation through an analytical benchmark and various two-dimensional simulations.
\end{abstract} 

\maketitle

\section*{Introduction}
Relativistic magnetohydrodynamics (MHD) provides a universal framework for describing the dynamics of conducting fluids in high-energy and astrophysical environments, ranging from quark–gluon plasmas to accretion flows around compact objects \cite{Jaiswal_2016,Hernandez_2017}. In many of these systems, dissipative transport coefficients such as viscosity, thermal conduction, and resistivity play a crucial role in determining the evolution of the fluid and the structure of magnetic fields. At the same time, incorporating dissipation in a relativistic setting presents a long-standing conceptual challenge: naive first-order theories lead to parabolic equations that violate causality and give rise to instabilities at high velocities.

A traditional resolution of this problem is provided by Müller–Israel–Stewart-type theories \cite{ISRAEL1979341,Muller:1967zza}, which promote dissipative currents to independent dynamical variables in order to restore hyperbolicity. While successful, this approach comes at the cost of introducing additional degrees of freedom and a substantial increase in complexity, both conceptually and numerically. More recently, a different route has emerged through the formulation of Bemfica–Disconzi–Noronha–Kovtun (BDNK) theory \cite{Kovtun_2019,PhysRevX.12.021044,Armas_2022,Hoult2020}, in which causality and stability are restored by including higher-derivative corrections to the conserved currents, without enlarging the set of dynamical variables. This framework has opened the possibility of constructing minimal, first-order relativistic theories that remain well-posed \footnote{Another potential resolution to the stability problem in relativistic hydrodynamics was proposed in \cite{Armas:2020mpr} and consists in recasting the fluid equations in an approximate Lorentzian form and in specific choice of hydrodynamic frame - the \quotes{density frame}.}.

Numerical tests of BDNK hydrodynamics have so far been carried out mostly in highly symmetric settings in flat backgrounds and without propagating magnetic fields. The earliest simulations considered effectively $1+1$-dimensional conformal flows, including smooth Gaussian pulses, shock tubes and steady shock-wave profiles \cite{Pandya_2021}. These were followed by conservative finite-volume BDNK schemes tested in $1+1$ and $2+1$ dimensions on Gaussian conservation tests, shock tubes, oblique shocks, viscous rotors, Kelvin--Helmholtz instabilities and inviscid-limit problems \cite{PandyaMostEtAl2022}, and by non-conformal/ideal-gas extensions involving $0+1$ Bjorken flow, planar shocks, relaxation tests and heat-flow problems \cite{Pandya:2022sff}. Other studies include $2+1$-dimensional comparisons against holographic microscopic evolution for quark-gluon-plasma-like flows \cite{Bantilan:2022ech}, $1+1$-dimensional frame-robustness tests using sinusoidal perturbations, Gaussian pulses and shock waves \cite{Bea:2023rru}, kinetic-theory/Bjorken-flow attractor studies \cite{Rocha:2022ind,Rocha:2023hts}, density-frame comparisons including BDNK runs \cite{Bhambure:2024axa}, radial-flow phenomenology for heavy-ion data \cite{Bea:2025eov}, constraint-propagation tests \cite{Fantini:2025gnm}, and recent flux-conservative shock-regularisation tests \cite{Clarisse:2025lli}. Together, these works demonstrate that BDNK schemes offer a reliable numerical approach to modelling dissipative effects in relativistic hydrodynamics.
\newline 
At first sight, BDNK theory does not appear to play the same role in magnetohydrodynamics (MHD) as it does in ordinary dissipative hydrodynamics. In conventional dissipative, ergo \quotes{resistive} relativistic MHD, Ampère's law gives the electric field its own retarded evolution, which renders the traditional formulation causal, but also introduces stiff relaxation that can make numerical evolution challenging \cite{Palenzuela_2009,PhysRevD.107.056003}. The one-form formulation of MHD offers a way to remove this stiffness by replacing Ampère's law with a conservation law for magnetic flux \cite{schubring,Grozdanov_2017,lier2025resistiverelativisticmagnetohydrodynamicsamperes,Wright_2019,hoult2024}. In doing so, however, the first-order resistive dynamics becomes parabolic. It is precisely at this stage that a BDNK-type completion becomes important also for the magnetic field evolution.  
\newline 
Despite substantial theoretical progress on one-form MHD, this step is nontrivial: while the structure of BDNK-type corrections for MHD is well understood at the level of continuum theory, it is far from obvious how to incorporate them into a stable and efficient numerical scheme in the presence of coupled dissipative processes and dynamical magnetic fields.
\newline 
Building on Ref.~\cite{lier2025resistiverelativisticmagnetohydrodynamicsamperes}, we address this gap by constructing and implementing a numerical scheme that combines viscous and resistive dissipation for relativistic MHD within the BDNK framework.  We show how the BDNK corrections to the conserved currents can be incorporated in a shock-capturing finite-volume scheme in a way that remains stable and computationally efficient.  Furthermore, we demonstrate how evolving the BDNK-type equations completely alleviates the need for non-linear primitive-variable recovery procedures. With a minimal amount of BDNK corrections, one obtains equations with a telegrapher-type structure associated with causal diffusion \cite{Cattaneo1948,Koide2007}. It turns out, however, that in order to guarantee the hyperbolic evolution of all diffusive variables, a fourth BDNK term must be added that does not have a simple telegrapher-type equation interpretation.

We validate our approach using a series of one- and two-dimensional tests, including an analytical benchmark, shock tubes, and configurations with strong spatial gradients. These results demonstrate that dissipative relativistic MHD in the BDNK framework can be implemented in a numerically robust way. Our work thus provides a concrete bridge between recent theoretical developments in relativistic hydrodynamics and their application to realistic simulations of magnetized relativistic flows.

\section{Equations of motion}
In this work, we consider a relativistic magnetohydrodynamic plasma. For simplicity, we focus on the ultra-relativistic limit where mass density does not play a role. To account for the dynamical magnetic field, we follow the \textit{one-form} approach \cite{schubring,Grozdanov_2017,Armas20201,Armas:2018atq}, which treats magnetic field lines as conserved string-like objects of codimension two. This means that, in the absence of external currents, the conservation laws are given by 
\begin{align} \label{eq:conservationlaws}
    \partial_{\nu} T^{\mu \nu }  =  0 ~~ , \quad
    \partial_{\nu} J^{\mu \nu }   =  0 ~~ , 
    \end{align}
where $T^{\mu \nu }$ is the stress-energy tensor and $J^{\mu \nu }$ is the dual field strength tensor.

Plasmas tend to be characterized by a large conductivity which screens the electric field, allowing it to be eliminated with the relation $\mathbf{E}=\mathbf{B}\times \mathbf{v}$, 
where $v^i$ is the fluid velocity in units of the speed of light, $E_i = F_{it}$ is the electric field, and $B^{i}= \frac{1}{2} \epsilon^{ijk}F_{jk}$ is the magnetic field.  
The resulting system of conservation laws is known as ``ideal MHD''. For ideal MHD, magnetic field lines are strictly topological in the sense that their connectivity is frozen and they only move with the flow \cite{GoedbloedPoedts2004}. We consider the simplest situation in which the magnetic field decouples from the equation of state such that the total thermodynamic pressure can be written as $P(T,b^2) = p(T) + b^2/2$ and the constitutive relations of ideal MHD become
\begin{subequations} \label{eq:idealconstitutiveequations2}
\begin{align}
    T_{(0)}^{\mu \nu } 
    & = \Big( \epsilon + p + b^2\Big) u^{\mu } u^{\nu} 
    +  \lb p + \half b^2\rb g^{\mu\nu}  
    - b^{\mu } b^{\nu }, \\   \label{eq:Jcurrent1}
    J_{(0)}^{\mu\nu} &= 2  u^{[\mu} b^{\nu] } ~~ .
    \end{align}
\end{subequations}
Here $u^\mu = \Gamma(1,{\bf v})$ is the fluid four-velocity, with the Lorentz factor $\Gamma=1/\sqrt{1-{\bf v}^2}$, and $b^\mu = \Gamma({\bf B}\cdot{\bf v},{\bf B}-{\bf v}\times{\bf E})$ is the magnetic field in the fluid rest frame with $b^2=b^\mu b_\mu$. Note that $b^\mu u_\mu=0$ and $u^\mu u_\mu=-1$. Furthermore, $\epsilon$ is the total fluid energy density and $p$ is the fluid pressure, which obey the following thermodynamic relations
\begin{align} \label{eq:thermodynamicrelations}
    \epsilon + p 
    =  Ts ~~, \qquad
    \df p =  s\, \df T  ~~ , 
\end{align}
where $T$ is the temperature and $s$ is the entropy density.
We close the equations with the ultrarelativistic equation of state
\begin{align}\label{eq:EoS}
    \epsilon =3  p  ~~. 
\end{align}
Let us now consider the first-order corrections to the ideal currents specified in \eqref{eq:idealconstitutiveequations2}, i.e. 
\begin{subequations} \label{eq:idealconstitutiveequations1}
\begin{align}
    T^{\mu \nu } 
    & =   T_{(0)}^{\mu \nu }  +A_{(1)} u^\mu u^\nu  + 2 u^{(\mu}  Q_{(1)}^{ \nu ) }  + \Pi_{(1)}^{\mu \nu } + \mathcal{O} ( \partial^2 ) , \\   \label{eq:Jcurrent12}
   J^{\mu \nu } 
    & =   J_{(0)}^{\mu \nu }  + 2 u^{[\mu} N_{(1)}^{\nu]}  + S_{(1)}^{\mu\nu}  + \mathcal{O} ( \partial^2 ) ~~ .  
    \end{align}
\end{subequations}
We consider all the dissipative corrections that are on the diagonal of the Onsager matrix for $T_{(1)}^{\mu \nu }$ and $J_{(1)}^{\mu \nu } $, which means we account for thermal conductivity, viscosity and resistivity, entering the currents as
    \begin{subequations}    \label{eq:firstorderequations}
        \begin{align}
         Q_{(1)}^{ \mu}  & =  - \sigma P^{\mu \rho }  \left(     u^{\nu}  \partial_{\nu} u_{\rho}  +    \partial_\rho\ln T  \right)   \\ 
         \Pi_{(1)}^{\mu \nu }   &  =   - 2  \eta   P^{\mu \rho }  P^{\nu \sigma } \partial_{ ( \rho}  u_{\sigma ) } -   \left(\zeta - \frac{2}{3} \eta  \right)   P^{\mu \nu }  P^{\rho \sigma } \partial_{  \rho}  u_{\sigma  } ~~ ,   \\ 
        S_{(1)}^{\mu  \nu}   &  =  - \left( 
    2 r_{\perp }  \mathbb{B}^{ \rho  [ \mu } \hat b^{\nu ] } \hat  b^{\sigma}   
    + r_{\parallel }  \mathbb{B}^{\mu \rho }  \mathbb{B}^{\nu \sigma } \right) 
    2T\partial_{[ \rho }\!\left(  \frac{b_{\sigma ] } }{T} \right)  ~~  ,     \end{align} 
 \end{subequations}
 where we introduced the tensors
 \begin{align*}
     P^{\mu \nu } = \eta^{\mu \nu } + u^{\mu } u^{\nu}   , ~~      \mathbb{B}^{\mu \nu } =  P^{\mu \nu }  -  \hat{b}^{\mu } \hat{b}^{\nu}    , ~~ \hat{b}^{\mu}  = b^{\mu} / \sqrt{b^2 } ~~ .  
 \end{align*}
 Note that we only considered the possibility of anisotropy for resistivity, whereas viscosity and thermal conductivity are treated as isotropic for simplicity. Our choice of frame allows for taking $N_{(1)}^{\nu } = A_{(1)}  =0 $. In addition, the second law of thermodynamics requires that 
 \begin{align}
    \eta, \zeta, r_{  \parallel }  \geq  0 ~~ ,~~r_\perp+\frac{b^2}{(\epsilon+p+b^2)^2}\sigma\ge 0~~. 
 \end{align}

\section{Stability and causality}
The first-order corrections of \eqref{eq:idealconstitutiveequations1} make the equations of motion parabolic. Parabolicity induces a violation of causality and thereby leads to instabilities when the fluid velocity approaches the speed of light \cite{HISCOCK1983466,hiscock,PhysRevX.12.041001}. One possible solution to this problem is to follow the M\"uller-Israel-Stewart (MIS) prescription~\cite{ISRAEL1979341,Muller:1967zza} and add new gapped tensor fields to regulate the instabilities \cite{Chandra:2015iza, Cordeiro:2023ljz,Molnr2009,PhysRevD.104.103028}. We instead implement the recently discovered Bemfica-Disconzi-Noronha-Kovtun (BDNK) prescription \cite{Kovtun_2019,PhysRevX.12.021044,Armas_2022,hoult2024,Hoult2020} by performing a zeroth order on-shell transformation \cite{hkzq-b2ph} to \eqref{eq:idealconstitutiveequations1} that renders Eq.~\eqref{eq:conservationlaws} causal. Additionally, the BDNK description allows us to set up a numerical scheme that efficiently performs primitive-variable recovery in the presence of dissipation \cite{PandyaMostEtAl2022,Pandya:2022sff}. We choose
        \begin{align}  \label{eq:BDNKterms}
\begin{split}
            A_{(1) \text{BDNK}}    & =  - \tau_{\epsilon} u_{\mu}  \partial_{\nu} T^{\mu \nu }_{(0) }    , ~~        Q_{(1) \text{BDNK}}^{ \mu}   =   \tau_{u} P^{\mu }_{\rho }  \partial_{\nu} T^{\nu \rho }_{(0) } ,   \\ 
         \Pi_{(1) \text{BDNK}}^{\mu \nu }   &  =   - \tau_{X}   P^{\mu \nu } u_{\rho}  \partial_{\sigma} T^{\rho \sigma  }_{(0)  }   ,   ~~  
        N_{(1) \text{BDNK}}^{\mu }     = -  \tau_{b} \partial_{\nu} J^{\mu \nu }_{(0) }     ,
        \end{split}
\end{align}
 whereas we take $    S_{(1) \text{BDNK}}^{\mu  \nu}  =0 $. In App.~\ref{eq:connecting}, we connect the coefficients introduced in \eqref{eq:firstorderequations} and \eqref{eq:BDNKterms} to those of the more general magnetohydrodynamic model of Ref.~\cite{Armas_2022}.
 
 The central role of these BDNK terms is to turn parabolic equations into telegrapher-type equations, which are hyperbolic. To understand how that works, let us first take $\tau_X :=0 $ and write out Eqs.~\eqref{eq:conservationlaws} with \eqref{eq:idealconstitutiveequations1}, \eqref{eq:firstorderequations} and \eqref{eq:BDNKterms} in the rest frame, yielding
 \begin{subequations} \label{eq:telegrapherequations}
    \begin{align}
    \label{eq:energytelegrapher}
&    \dot{\epsilon} -  ( D_{\epsilon}  - \tau_u   )   \Delta  \epsilon / 3    +   \tau_{\epsilon } \ddot{\epsilon}     + ...   =0    ~~ , \\
 &      \dot{\mathbf{u}  } -   D_u   (  \Delta      +      \nabla \nabla \cdot  )\mathbf{u}        +   ( \tau_u   - D_{\epsilon } )  \ddot{\mathbf{u}  }  +... = 0 ~~ ,  \\
&  \dot{\mathbf{b}  } - r^{\prime}_b  \Delta    \mathbf{b}   +  \tau_{b}  \ddot{\mathbf{b}  }   +  ...  =0  ~~ ,   \label{eq:bterms}
\end{align}
\end{subequations}
where $\Delta $ is the Laplacian operator and $...$ stands for terms nonlinear in the variables $\epsilon$, $\mathbf{u}$ and $\mathbf{b}$ or terms that couple the three equations. Furthermore, we took for simplicity
\begin{align}  \label{eq:simplificationsenergymommentum}
    \sigma =  w D_{\epsilon}  , ~~   \eta = w D_u  , ~~   \zeta  =  \frac{2 w D_u }{3}   , 
    \end{align}
    and 
    \begin{align}
    r_{\parallel } := r_{\perp} := r^{\prime}_b ~~,
\end{align}
with $D_u$, $D_{\epsilon}$ and $r^{\prime}_b$ constants. Note that in the rest frame, $\mathbf{b}$ satisfies the magnetic Gauss law 
\begin{align}
 \nabla \cdot \mathbf{b} =0 ~~ .      
\end{align}
It follows from \eqref{eq:telegrapherequations} that the BDNK terms $\tau_{\epsilon}, \tau_u $ and $\tau_{b}$ turn the parabolic equations for $\epsilon$, $\mathbf{u} $ and $\vct b$ into telegrapher-type equations, which are equations that have a structure that is suited for introducing dissipation in a causal way \cite{Cattaneo1948,Koide2007}. When evolution is causal, stability in one frame implies stability in any frame \cite{PhysRevX.12.041001,PhysRevLett.128.010606}. It follows from \eqref{eq:bterms} that when one has $\tau_b  > r^{\prime}_b $, the front velocity of $\mathbf{b}$ is subluminal and its evolution is therefore causal \cite{Krotscheck1978,PhysRevD.109.046018}. When $D_{\epsilon}  > \tau_u$, the accelerative contribution to thermal diffusion makes the equation for $\vct u $ elliptic \cite{PhysRevD.35.3723}. Conversely, when $D_{\epsilon} < \tau_u$, it follows from \eqref{eq:energytelegrapher} that the equation for $\epsilon$ becomes elliptic, which means that these three BDNK coefficients alone do not suffice to make the full set of equations hyperbolic. To resolve this issue, we introduce the fourth BDNK term $\tau_X$. To see its effect, we consider the linearized modes $\omega (k)$ of \eqref{eq:conservationlaws} for a uniform background, i.e.
\begin{align}
   \epsilon =  \epsilon_0   + \delta  \epsilon , ~~     \vct u =   \vct u_0   + \delta   \vct u 
 , ~~     \vct b =   \vct b_0   + \delta   \vct b ~~ .  
\end{align}
Since this issue pertains to $ \vct u$ and $ \epsilon$, we take $ \vct b_0 =0 $ for this analysis, which causes the dynamics of the magnetic field to decouple. We furthermore consider the background fluid velocity to be in the rest frame so that we have $\vct u_0 =0 $. Defining the front velocity $W$ as 
\begin{align}
    W = \lim_{k \rightarrow \infty }  \frac{\omega (k)}{k} ~~ , 
\end{align}
we then find for the sound sector
\begin{widetext}
\begin{align}
    W^2 =
    \frac{
    6 D_u \tau_\epsilon
    + (3\tau_X+\tau_\epsilon)(\tau_u-D_\epsilon)
    \pm
    \sqrt{
    12\tau_\epsilon(2D_u-\tau_X)(\tau_u-D_\epsilon)^2
    +
    \left[
    6D_u\tau_\epsilon
    +
    (\tau_u-D_\epsilon)(3\tau_X+\tau_\epsilon)
    \right]^2
    }
    }{
    6\tau_\epsilon(\tau_u-D_\epsilon)
    } ~~ .
\end{align}
\label{eq:Ws}
\end{widetext}
This expression shows that the  front velocities are real provided
\(\tau_u>D_\epsilon\), \(\tau_\epsilon>0\), and
\(\tau_X\geq 2D_u\). Subluminality imposes an additional upper bound on
\(\tau_X\). In particular, the larger scalar branch satisfies \(W^2\leq 1\)
if
\begin{align}
    2D_u
    \leq
    \tau_X
    \leq
    \tau_\epsilon
    -
    D_u
    -
    \frac{3D_u\tau_\epsilon}{\tau_u-D_\epsilon} .
\end{align}
We also note that requiring stability around this same equilibrium configuration of velocity and magnetic fields at arbitrary $k$ does not result in additional conditions to those arising from causality.

\vspace{0.5em}
\noindent
\textit{Numerical scheme---}%
We now outline a numerical scheme to solve \eqref{eq:conservationlaws} with the constitutive equations described by Eqs.~\eqref{eq:conservationlaws} with \eqref{eq:idealconstitutiveequations1}, \eqref{eq:firstorderequations} and \eqref{eq:BDNKterms}. We again fix the diffusive coefficients pertaining to energy and momentum as in \eqref{eq:simplificationsenergymommentum}, whereas for resistivity we take
\begin{align} \label{eq:condition}
  \frac{w + b^2}{w}   r_{\perp} := r_{\parallel}  := r_b  ~~ , 
\end{align}
with $r_b$ a constant. Eq.~ \eqref{eq:condition} allows one to straightforwardly connect the linearized modes of one-form MHD to those of traditional resistive MHD \cite{Armas20201,lier2025resistiverelativisticmagnetohydrodynamicsamperes}. We also take for simplicity $\tau_{\epsilon} := 2 \tau_{u} $, so that we have 
\begin{align}
   \frac{1}{2}   A_{(1) \text{BDNK}}  u^{\mu}  +       Q_{(1) \text{BDNK}}^{ \mu} :  =   \tau_{u}  \partial_{\nu} T^{\nu \mu }_{(0) } ~~  .  
\end{align}  
Our implementation is added as a physics module to \texttt{BHAC} \citep{PorthOlivares2017,OlivaresPorthEtAl2019} which offers various numerical schemes to solve conservation laws on arbitrary background metrics. To make the equations amenable to numerical integration, we project them onto space-like hypersurfaces. This allows us to decompose the currents into conserved variables \(U_I\) and fluxes \(\mathbf{F}_I\), where \(I=1,\ldots,7\) labels the seven hydrodynamic components considered in this work.
 Simultaneously with the conserved variables, we evolve primitive variables $ P_I $ with sources $S_I $. That is, we evolve as
\begin{subequations} \label{eq:fluxequation} 
    \begin{align}
    \partial_t{U}_I +   \nabla \cdot  \mathbf{F}_I = 0  , ~~    \partial_t{P}_I  = S_I  ~~  , 
\end{align}
with 
\begin{align}
 U_I = \begin{pmatrix}
        J^{tl} \\ 
    T^{tl}\\
    T^{tt}     \end{pmatrix}, ~   
  \mathbf{F}_I = 
  \begin{pmatrix}
    \mathbf{J}^{l}  \\ 
   \vct T^{l} \\
   \vct T^{t} \\
  \end{pmatrix}, ~  P_I = \begin{pmatrix}
    b^l  \\ 
  u^l  \\ 
  \epsilon     \end{pmatrix} , ~ 
  S_I = 
  \begin{pmatrix}
    \dot{b}^l \\
    \dot{u}^l \\
    \dot{\epsilon} \\
  \end{pmatrix} ,   
\end{align}
\end{subequations}
where $l$ represents the spatial index. Equation \eqref{eq:fluxequation} is implemented in \texttt{BHAC} using a finite-volume discretization. We adopt a second-order total variation diminishing time-stepper and spatial reconstruction techniques.

As in our previous work on purely resistive one-form MHD \cite{lier2025resistiverelativisticmagnetohydrodynamicsamperes}, gradients appearing in the dissipative fluxes $\mathbf{F}_I$ are evaluated by second-order central differencing of the reconstructed interface states, yielding left-biased and right-biased fluxes. From these, we compute upwinded numerical fluxes using an approximate Riemann solver, see e.g. Sec. 2.9 of \cite{PorthOlivares2017} for details.
Similarly, we use unlimited second-order central differencing to obtain gradients for the source term $ S_I $. For simplicity, the characteristic velocities are set to the speed of light which provides an upper limit to the true characteristic velocities (at the cost of a slightly more diffusive algorithm).  
To ensure that $\partial_i J^{it}=0$ is maintained to machine precision, we utilize the existing schemes in \texttt{BHAC} such as the cell-centered flux-constrained transport algorithm (FCT) described by \cite{Toth2000} and the staggered constrained transport scheme due to \cite{BalsaraSpicer1999,PorthOlivares2017}. 
\newline 
To close the finite-volume update, the primitive-variable time derivatives
$S_I$ have to be recovered. The conserved variables depend algebraically and
linearly on $S_I$ through the first-order contributions given in \eqref{eq:idealconstitutiveequations1}. At fixed primitive variables $P_I$ and fixed spatial
gradients, we may therefore write
\begin{subequations}
\label{eq:primitive_recovery}
\begin{align}
   U_I(S_K) = U_I(0) + \sum_J M_{IJ} S_J ,
   \label{eq:conserved_linear_in_sources}
\end{align}
where $U_I(0)$ denotes the conserved variables evaluated with $S_I=0$.
The primitive source terms are then obtained by the local matrix inversion
\begin{align}
   S_I =
   \sum_J (M^{-1})_{IJ}
   \left[ U'_J - U_J(0) \right] .
   \label{eq:source_recovery}
\end{align}
Here $U'_J $ is the conserved state supplied by the finite-volume update.
The matrix $M_{IJ}$ is computed locally by evaluating the constitutive
relations on unit vectors $e_J$,
\begin{align}
   M_{IJ} = U_I(e_J) - U_I(0) ~.
   \label{eq:source_matrix}
\end{align}
\end{subequations}
This procedure turns the primitive-variable recovery into an
algebraic problem. The BDNK coefficients given in \eqref{eq:BDNKterms} are essential in this step:
they guarantee that the matrix $M_{IJ}$ is non-degenerate \cite{PandyaMostEtAl2022,Pandya:2022sff}.

It is important to stress that the algorithm remains fully conservative: the conserved variables $U_I$ are fully determined by their numerical fluxes and are conserved to machine precision in the finite volume discretization.  
Furthermore, we emphasize that the evolution equations for the primitive variables in \eqref{eq:fluxequation} come from the appearance of time-derivatives in the gradient expansion for the conserved variables and not from an additionally postulated non-conservative formulation of the dynamics. Thus our algorithm is fundamentally different from evolving auxiliary non-conservative equations of motion as in, for example, 'dual energy' formulations \cite{AnninosFragile2005}.  Such an approach was suggested in \cite{Pandya:2022sff} as a backup solution in regions where the physical dissipation becomes smaller than the numerical one.  For the test cases presented in this paper, however, no such backup strategy is required.  

\section{Numerical evolution}
In this section, we discuss various benchmark problems with increasing complexity.  An overview of the performed simulations is given in Table \ref{tab:simulations}.  

\begin{table*}
 \caption{Overview of the performed simulations and corresponding parameter choices. }
 \label{tab:simulations}
 \begin{tabular*}{\textwidth}{p{3.5cm}p{1.5cm}cccccc}
  \hline 
  Type & ID & $D_u$ & $D_\epsilon$ & $r_b$ & $\tau_u$ & $\tau_X$ & $\tau_b$ \\
  \hline
  \multirow[t]{2}{3.5cm}{\parbox[t]{3.5cm}{1D Shock tubes}} &
  ST-\{a-i\} & $\{10^{-4},10^{-3},10^{-2}\}$ & $D_u$  & $\{10^{-4},10^{-3},10^{-2}\}$ & $2\times D_u$ & $2\times D_u$ & $2\times r_b$ \\
  & ST-j & $10^{-2}$ & $10^{-2}$ & $10^{-2}$ & $2\times10^{-2}$ & $4\times10^{-2}$ & $2\times10^{-2}$\\
  \hline
  \multirow[t]{4}{3.5cm}{\parbox[t]{3.5cm}{2D Kelvin-Helmholtz}} &
  KH-a & $10^{-4}$ & $5\times 10^{-5}$ & $10^{-4}$ & $10^{-3}$ & $5\times 10^{-4}$ & $5\times 10^{-4}$ \\
  &
  KH-b & $10^{-4}$ & $5\times 10^{-5}$ & $10^{-3}$ & $10^{-3}$ & $5\times 10^{-4}$ & $5\times 10^{-3}$ \\
  &
  KH-c & $10^{-3}$ & $5\times 10^{-4}$ & $10^{-4}$ & $8\times 10^{-3}$ & $5\times 10^{-3}$ & $5\times 10^{-4}$ \\
  &
  KH-d & $10^{-3}$ & $5\times 10^{-4}$ & $10^{-3}$ & $8\times 10^{-3}$ & $5\times 10^{-3}$ & $5\times 10^{-3}$ \\
    \hline
\multirow[t]{5}{3.5cm}{\parbox[t]{3.5cm}{2D Orszag--Tang}} 
& OT-a
& $10^{-2}$
& $2\times10^{-3}$
& $10^{-2}$
& $2\times10^{-1}$
& $2\times10^{-1}$
& $8\times10^{-2}$ 
\\
& OT-a-tx0
& $10^{-2}$
& $2\times10^{-3}$
& $10^{-2}$
& $2\times10^{-1}$
& $0$
& $8\times10^{-2}$
\\
& OT-b
& $10^{-1}$
& $2\times10^{-2}$
& $10^{-2}$
& $2\times10^{0}$
& $2\times10^{0}$
& $8\times10^{-2}$ 
\\
& OT-c
& $10^{-2}$
& $2\times10^{-3}$
& $10^{-1}$
& $2\times10^{-1}$
& $2\times10^{-1}$
& $8\times10^{-1}$ 
\\
& OT-d
& $10^{-1}$
& $2\times10^{-2}$
& $10^{-1}$
& $2\times10^{0}$
& $2\times10^{0}$
& $8\times10^{-1}$
\\
  \hline
 \multirow[t]{4}{3.5cm}{\parbox[t]{3.5cm}{2D Harris sheet}} &
 HS     & $10^{-2}$& $5\times 10^{-3}$ & $10^{-2}$ & $10^{-1}$& $5 \times 10^{-2}$& $ 10^{-1}$ \\
 \hline
  \end{tabular*}
\end{table*}

\subsection{Correctness and convergence}
\begin{figure}
    \centering
    \includegraphics[width=0.9\linewidth]{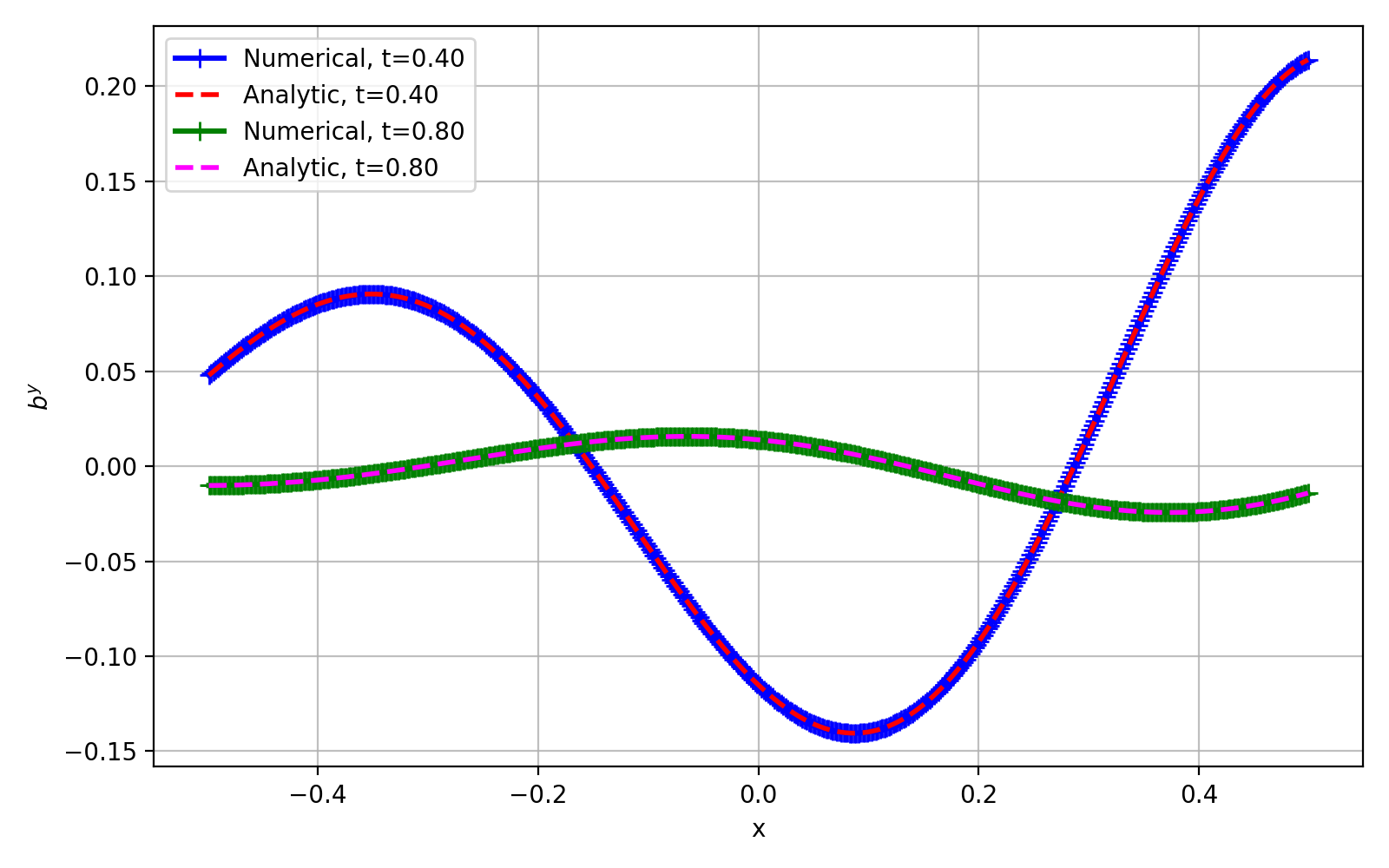}
    \caption{Comparison of the numerical output with the analytical solution of \eqref{eq:bthing}.}
    \label{fig:placeholder12223333}
\end{figure}

\begin{figure}
    \centering
    \includegraphics[width=0.9\linewidth]{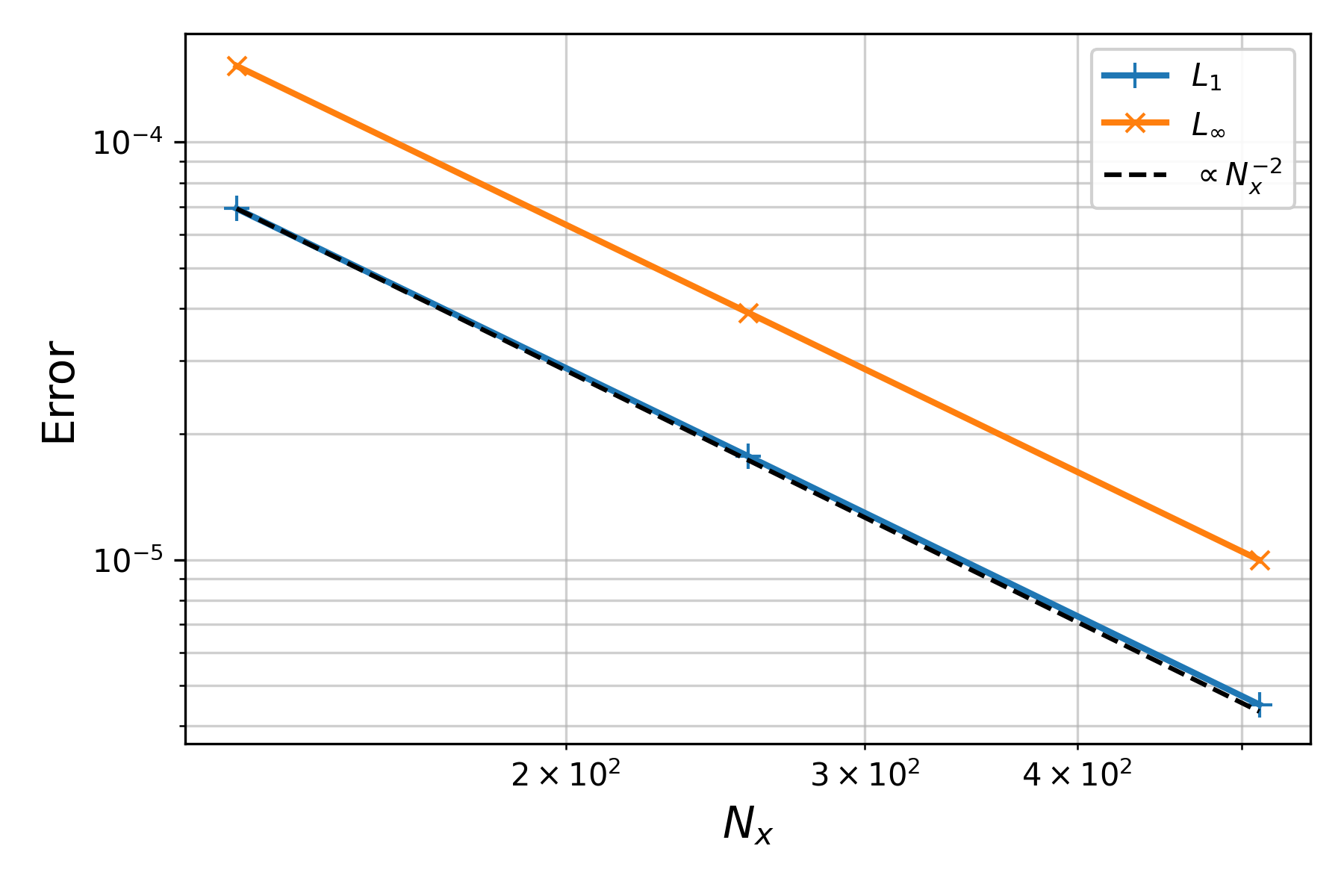}
    \caption{Convergence for the boosted telegraph solution of \eqref{eq:bthing} for simulation at $t=0.8$. }
    \label{fig:convergence}
\end{figure}

To verify correct implementation, we first consider an analytical benchmark.  Since finding analytic solutions for the fully coupled viscoresistive set is difficult, we here settle for a simplified test that was previously discussed in \cite{lier2025resistiverelativisticmagnetohydrodynamicsamperes}.  We write the magnetic field evolution equation as the telegrapher-type equation that is shown in \eqref{eq:bterms}, taking all other primitive variables to be non-evolving. In the rest frame, we then have
\begin{subequations}  \label{eq:bthing}
    \begin{align}
  &b^y (t,x)
  = \exp\!\left(-\frac{t}{2\tau_b}\right)
    \sin\!\big(k x - \Theta_{b} t\big) ~~  , 
\end{align}
with $ \Theta_{b}^2
  = \frac{r'_b}{\tau_b}\,k^2
    - \frac{1}{4\tau_b^2} $. Boosting in the $x$-direction we can find a new solution given by
\begin{align}  
    b^{\prime y}   (t', x^{\prime } )   &  =   b^{ y}   ( \Gamma   t'  - u^x x^{\prime  },  \Gamma x^{\prime }  -   u^x  t'  ) ~~ . 
\end{align}
\end{subequations}
In order for this solution to be well-behaved for time-dependent non-periodic boundary conditions, we must explicitly turn off the evolution of the other primitive variables as well as take resistivity to be isotropic, i.e. $r_{\perp} := r_{\parallel} := r_b'  $ while in the remainder we take \eqref{eq:condition}. In Fig.~\ref{fig:placeholder12223333}, we show the overlap between the analytical and numerical results at different times. In Fig.~\ref{fig:convergence}, we show that the difference between the analytical and the numerical results converges to zero at the expected second-order rate when increasing the grid size.  While not covering the full set of implemented terms, we expect our implementation of the energy-momentum sector also to be correct since it is based on the same algebraic structure of \eqref{eq:primitive_recovery}.  

\subsection{1D shocktube tests}

Shock tubes are stringent tests for the stability of the proposed algorithm \citep[e.g.][]{BrioWu1988}.  For the dissipative scheme, the shocktube presents a formidable challenge since the initial condition is far out of equilibrium in the thermodynamic quantities (gradients are formally infinite).  
The initial condition simply consists of two states (left and right) which are separated by a discontinuity at $x=0$.  Similar to previous studies in relativistic MHD \cite{Palenzuela_2009, DionysopoulouAlic2013, Wright_2019,lier2025resistiverelativisticmagnetohydrodynamicsamperes}, we adopt the following initial states:
\begin{align}
    \begin{split}
        \left(p^L,J^{ty,L}\right) &= (1,1/2)~, \\
        \left(p^R,J^{ty,R}\right) &= (1/10,-1/2)\, ,
    \end{split}
\end{align}
while all other quantities are initialized with zero.  Note, however, that a direct comparison to relativistic MHD with mass density cannot be made the model used is ultra-relativistic.  
We scan over the parameters $r_{b}\in[10^{-4},10^{-3},10^{-2}]$ and $D_{u}=D_{\epsilon}\in[10^{-4},10^{-3},10^{-2}]$.  The BDNK terms are controlled by setting the relaxation times $\tau_u $, $\tau_{b}$ and $\tau_X$ to a value twice their corresponding dissipative parameters (except for the case $r_b=D_u=D_\epsilon=10^{-2}$ where we had to set $\tau_X=4\times 10^{-2}$ for numerical stability).  The simulation is performed in a 1D domain $x\in[-0.5,0.5]$ with a resolution of 1024 cells and a CFL parameter of 0.2.

\begin{figure*}
    \centering
    \includegraphics[width=\textwidth]{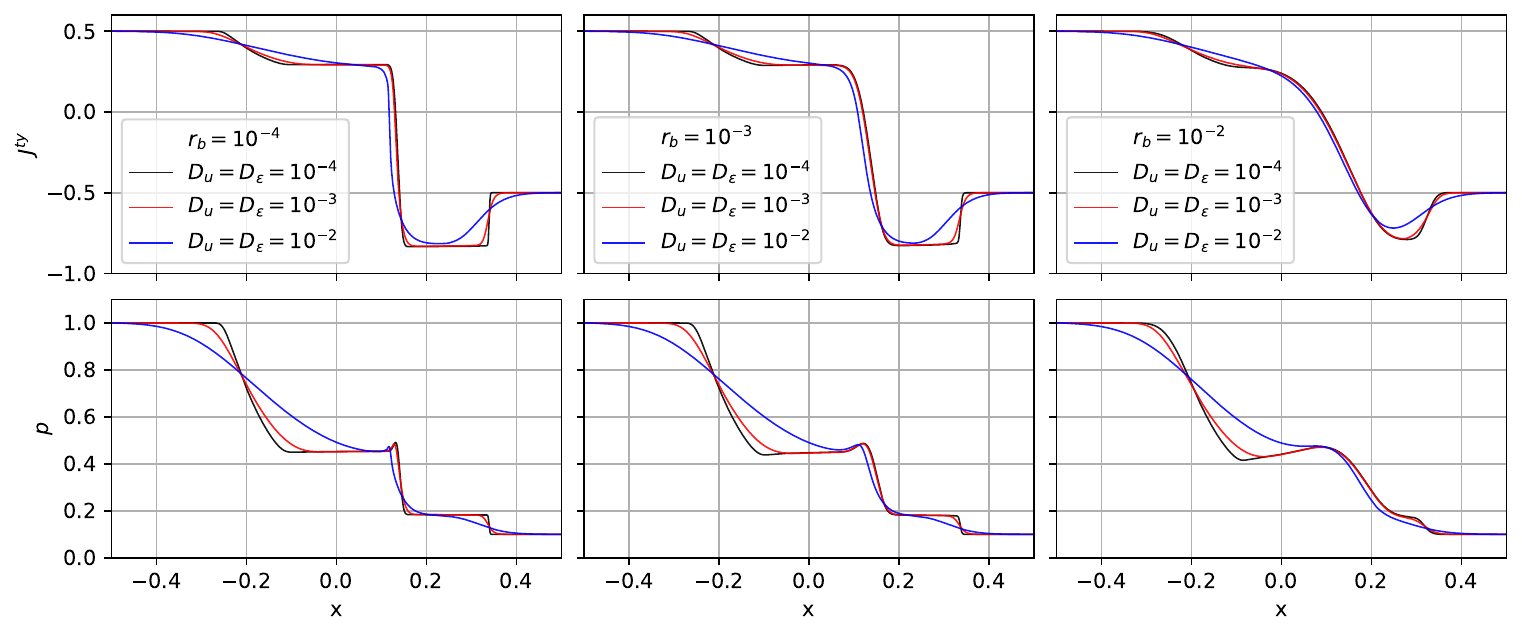}
    \caption{Shock tubes for a range of resistivities $r_b$ and momentum/energy diffusion parameters $D_{u}, D_{\epsilon}$.  The latter two are assumed equal in this test.  We show the out-of-plane magnetic field component $J^{ty}$ and the pressure $p=\epsilon/3$. The simulations span a large range of magnetic Prandtl number $D_u/r_b\in [0.01,100]$ and cover a factor of 100 in the (magnetic-) Reynolds number, clearly reproducing ideal and diffusion-dominated dynamics.  }
    \label{fig:BriuWu}
\end{figure*}

The solutions at time $t=0.4$ are shown in Figure \ref{fig:BriuWu}.  For small dissipative parameters, the solution closely resembles the known ideal MHD case \citep[e.g.][]{lier2025resistiverelativisticmagnetohydrodynamicsamperes}.  Increasing energy- and momentum diffusion has a strong impact on left-moving fast and rarefaction waves and on the right-moving fast wave.  The perpendicular shock at $x\simeq 0.15$ is less affected by momentum diffusion, in particular when resistivity dominates (see e.g. case $r_b=10^{-2}$).  Indeed, magnetic diffusion has the effect of smoothing out the solution most in regions of the strongest magnetic gradients which is reflected in the impact that resistivity has on the perpendicular shock.

\subsection{Viscoresistive Kelvin-Helmholtz instability}

The Kelvin-Helmholtz instability is a frequently encountered instability for subsonic shear flows.  It is well known that the tension from a magnetic field along the vortex sheet can suppress the instability \cite{Chandrasekhar1981-nb,Osmanov_2008,10.1093/mnras/198.4.1065}.  Furthermore, the instability can be avoided in the presence of viscous momentum transport.  
It is therefore an excellent way to investigate the effect of the different transport coefficients $D_{\epsilon}$, $D_{u}$ and $r_b$ in the ultra-relativistic case. The initial conditions we consider are given by
\begin{subequations}
    \label{eq:initialization}
\begin{align}  
u^x  &= 0.15 \left[
\tanh\!\left(\frac{y + 0.5}{0.05}\right)
- \tanh\!\left(\frac{y - 0.5}{0.05}\right)
- 1
\right], \\[1em]
u^y &= 10^{-2} \, \sin(2\pi x)
\sum_{\pm}
\exp\!\left(-\frac{(y \pm 0.5)^2}{0.04}\right)
 , \\[1em]
J^{tx} &= 0.08, \qquad
J^{ty} = 0, \qquad
J^{tz} = 0.8 \, .
\end{align}
\end{subequations}
The problem is hence doubly periodic and we adopt a domain size that is twice as large in the $y$-direction as it is in the $x$-direction, where $L_x=1$. The simulations are performed on a grid with $N_x \times N_y= 256\times 512$ cells and a CFL parameter of 0.2.  
For the purpose of visualization, we add a passively evolved tracer density 
$n$ that evolves according to
\begin{align}
    \partial_{\mu} ( n  u^{\mu} )  =0  ~~ ,
\end{align}
and it is initialized as
\begin{align}
    n &= 
1 + \tfrac{1}{2}
\left(
\tanh\!\left(\frac{y + 0.5}{0.05}\right)
- \tanh\!\left(\frac{y - 0.5}{0.05}\right)
\right)~.
\end{align}

\begin{figure*}
    \centering
\includegraphics[width=1\linewidth]{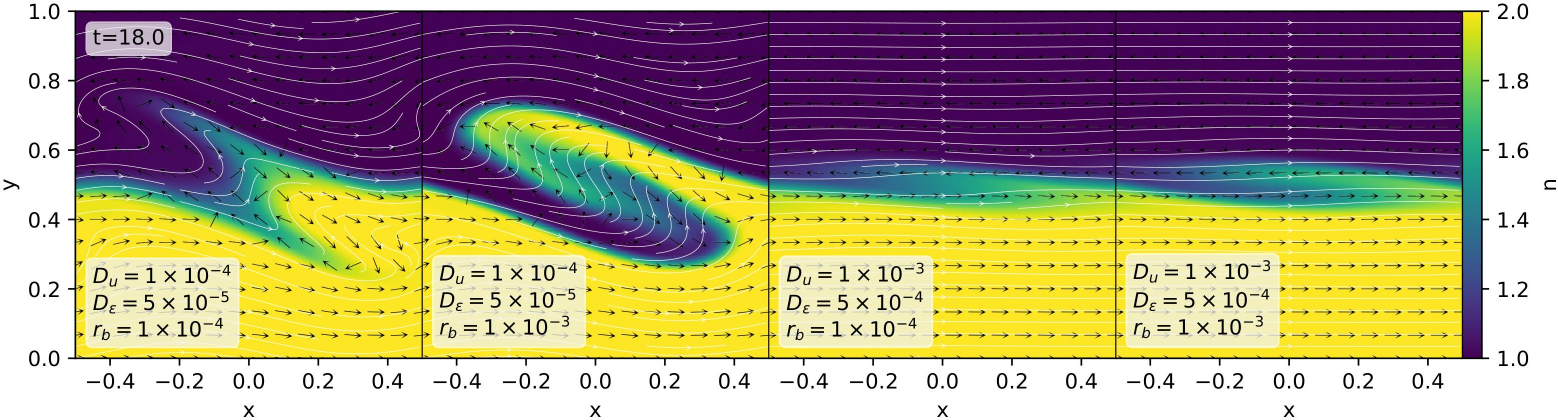}
    \caption{Color plot of late-time tracer density $n$ for the Kelvin-Helmholtz initialization of \eqref{eq:initialization} for four different values of $D_{u}$, $D_{\epsilon}$ and $r_b $. The black arrows represent fluid velocity $u^i$ whereas the white lines represent magnetic field lines $J^{t  i }  $.  See Table \ref{tab:simulations} for the adopted BDNK coefficients in each case.  }
    \label{fig:placeholder3}
\end{figure*}
\begin{figure*}
    \centering
\includegraphics[width=1\linewidth]{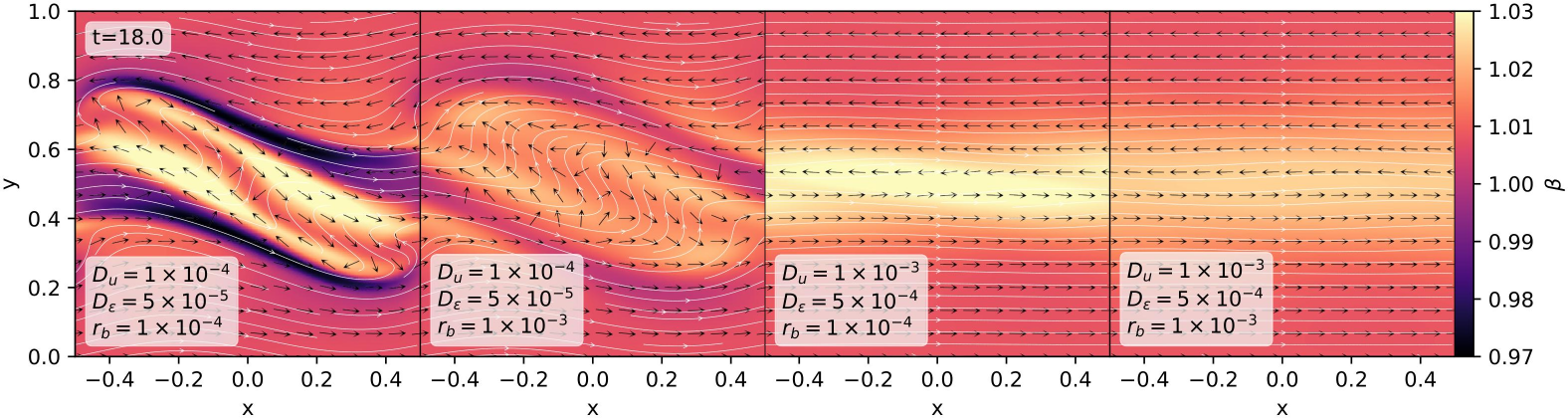}
    \caption{Color plot of the late-time $\beta$ parameter given by \eqref{eq:betacoefficient} for the same parameter choices as Fig.~\ref{fig:placeholder3}.}
    \label{fig:placeholder4}
\end{figure*}

The BDNK parameters are chosen to yield real and causal front velocities in the initial condition and we monitor possible violation of causality during runtime.  Quantitatively, for cases KH-a to KH-d as defined in Table \ref{tab:simulations}, the fastest initial front velocities in the computational domain are $0.9098, 0.9098, 0.9503, 0.9503$ and the maxima are $0.9139, 0.9102, 0.9503, 0.9503$ throughout the entire evolution.  

In Fig.~\ref{fig:placeholder3} the distribution of the tracer density is shown at time $t=18$ for a range of dissipative parameters (see Table \ref{tab:simulations}).  
We see that increasing $D_{u}$ and $D_{\epsilon}$ strongly undermines the Kelvin-Helmholtz instability. This can be intuitively understood: $D_{u}$ enables viscous momentum transport away from the interface which reduces the energy available to drive the instability.  Conversely, increasing $r_b$ enhances the instability as it reduces the coupling between the fluid velocity and the magnetic field lines, i.e. it relaxes Alfv\'en's theorem which holds in the ideal MHD case.  In Fig.~\ref{fig:placeholder4} we also plot the plasma-beta parameter given by
\begin{align} \label{eq:betacoefficient}
    \beta = \frac{2 p}{b^2} ~~ . 
\end{align}

\subsection{Orszag-Tang vortex}

\begin{figure*}
    \centering
    \includegraphics[width=1\linewidth]{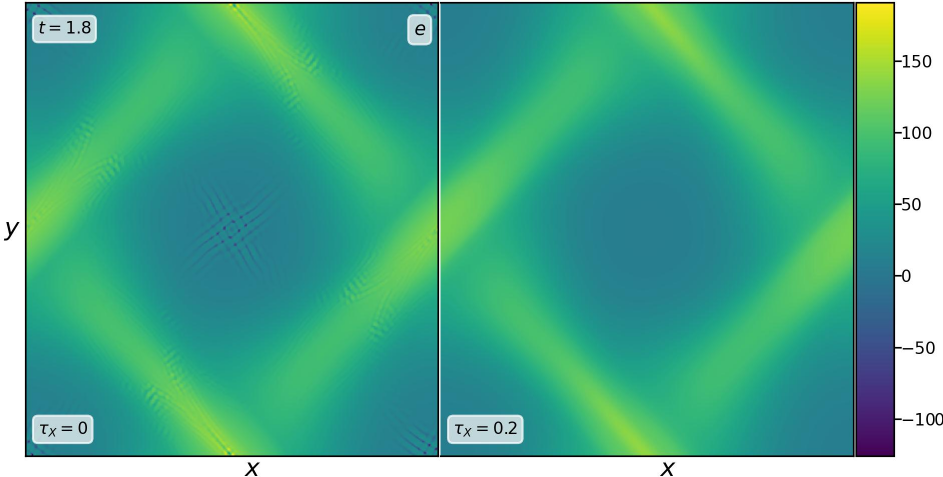}
    \caption{Comparison of lab frame energy density $e = T^{tt}$ at Orszag-Tang vortex evolution at early times for $\tau_X =0 $ and $\tau_X=0.2$, corresponding to cases OT-a-tx0 and OT-a in Table \ref{tab:simulations}.  }
    \label{fig:placeholder12123}
\end{figure*}

Another standard two-dimensional test is the Orszag-Tang vortex \citep{OrszagTang1979} where current sheets develop after an initial ideal evolution of MHD. The initial vector components of the Orszag-Tang vortex are given by
\begin{equation}   \label{eq:initializationOT}
    \begin{aligned} 
    \big(u^x, u^y\big)  
    &= \Big({-}\Gamma v_{\rm init} \sin(y), \Gamma v_{\rm init} \sin(x)\Big)~,\\
    \big(J^{tx},J^{ty}\big) 
    &= \Big({-}\sin(y), \sin(2x)\Big)\, ~,
\end{aligned}
\end{equation}
together with energy density $\epsilon =30 $. We choose $\Gamma  v_{\rm init}=0.8$. The comoving magnetic field is initialized under the assumption of negligible gradient terms such that $b^i = (J^{ti} + b^t u^i)/\Gamma$ and $b^t=J^{ti} u_i$. The doubly periodic computational domain is given by $x,y\in[0,2\pi]$ and resolved by $512^2$ cells.  This test is executed with a timestep corresponding to a Courant parameter of ${\rm CFL}=0.2$ and a $\partial_i J^{it}=0$-preserving FCT algorithm. We first show in Fig.~\ref{fig:placeholder12123} that when $\tau_X =0$, ripples arise in the lab frame energy density $e = T^{tt}$ which cause the code to crash. 

\begin{figure*}
    \centering
    \includegraphics[width=1\linewidth]{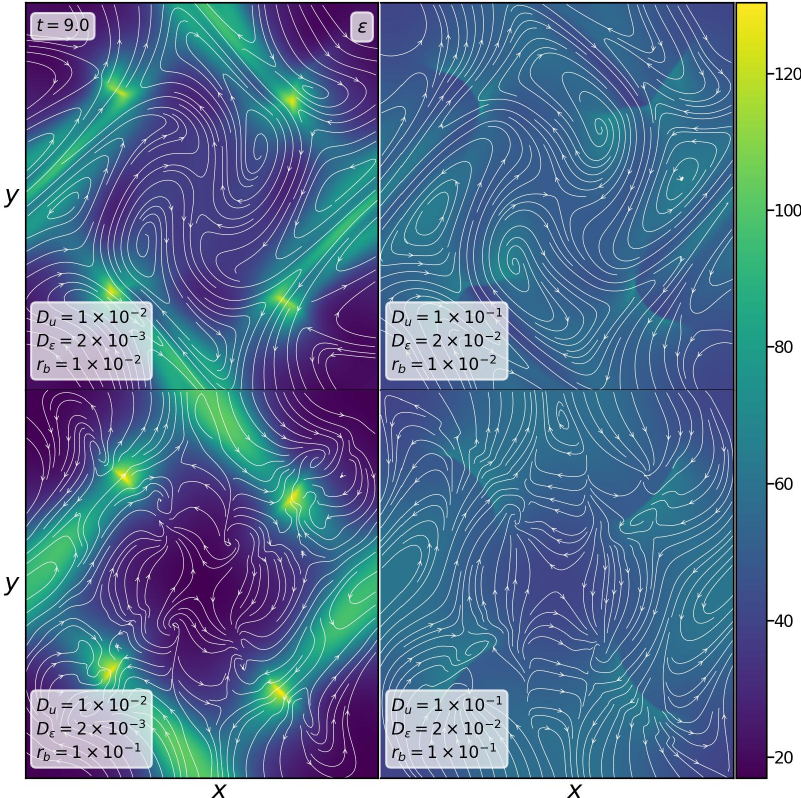}
    \caption{Color plot of late-time energy density $\epsilon$ initialized with an Orszag-Tang vortex of \eqref{eq:initializationOT} for different diffusion coefficients. We took $\tau_X = \tau_u = 20 D_u$ and $\tau_b = 8 r_b$. The white lines represent magnetic field lines $J^{t  i }  $.}
    \label{fig:placeholder123xx}
\end{figure*}

In this case, the maximal front velocity in the domain becomes $0.74439+0.18834 i$, corresponding to anti-diffusive behavior, consistent with \eqref{eq:energytelegrapher}.  Fig.~\ref{fig:placeholder12123} thus highlights the importance of this coefficient for stability. In the right-hand panel of Fig. ~\ref{fig:placeholder12123}, we have chosen $\tau_X =0.2$, resulting in an initial maximal front velocity of $0.86717$ which remains bounded and real throughout the entire evolution.  
Then, we show in Fig.~\ref{fig:placeholder123xx} the late-time energy density $\epsilon$ for different diffusive parameters. We see that increasing resistivity smooths the sharp ridge-like structures of the energy density induced by the Orszag-Tang vortex, whereas increasing $D_u$ and $D_\epsilon$ weakens the nonuniformity of energy density overall, leaving the creases intact.  

\subsection{Harris sheet}\label{sec:harris}

As a final two-dimensional test, we consider a perturbed double Harris
sheet. This setup is useful because it contains thin magnetic reversal
layers and therefore probes the behavior of the scheme in regions with
large magnetic gradients. The computational domain is periodic in the
$x$-direction and contains two current sheets located at $    y_1 = \frac{L_y}{4}$ and $  y_2 = \frac{3L_y}{4} $, where \(L_y\) is the vertical domain size. The initial fluid velocity is set to zero, i.e. $  u_{\text{init}}^x=u_{\text{init}}^y=u_{\text{init}}^z=0 $. The magnetic field is initialized from a vector potential with only a
nonzero \(z\)-component, \cite{Keppens}
\begin{widetext}
    \begin{align}
\begin{split}
 A_z ={}&
 - B_0 \left[
 y
 - \ell \log \cosh\!\left(\frac{y-y_1}{\ell}\right)
 + \ell \log \cosh\!\left(\frac{y_2-y}{\ell}\right)
 \right]
\\
&+ \psi_{\rm bot}
 \cos\!\left(k_x(x-x_0)\right)
 \cos\!\left(k_y(y-y_1)\right)
 \exp\!\left[-k_x(x-x_0)^2-k_y(y-y_1)^2\right]
\\
&- \psi_{\rm top}
 \cos\!\left(k_x(x-x_0)\right)
 \cos\!\left(k_y(y-y_2)\right)
 \exp\!\left[-k_x(x-x_0)^2-k_y(y-y_2)^2\right] ,
\end{split}
\label{eq:harris_vector_potential}
\end{align}
\end{widetext}
where \(x_0=L_x/2\), \(k_x=2\pi/L_x\), and \(k_y=2\pi/L_y\). The
unperturbed part of \eqref{eq:harris_vector_potential} gives the usual
magnetic reversal profile,
\begin{align}
\begin{split}
        J^{tx}
   &   =
    B_0\left[
    -1
    +\tanh\!\left(\frac{y-y_1}{\ell}\right)
    +\tanh\!\left(\frac{y_2-y}{\ell}\right)
    \right],  \\  
    J^{ty}  & =0 ,
    \end{split}
\end{align}
while the localized perturbations seed reconnection near the two sheets.
In the simulations shown below we take
\begin{equation}
    B_0  =1,\qquad
    \ell  =0.5,\qquad
    \psi_{\rm bot}  =\psi_{\rm top}=0.1 .
\end{equation}
The passive density and energy density are initialized as
\begin{align}
    D_{\text{init}} &=
    \rho_{\rm bg}
    + \rho_{\rm sh}\,
    \sech^2\!\left(\frac{y-y_1}{\ell}\right)
    + \rho_{\rm sh}\,
    \sech^2\!\left(\frac{y-y_2}{\ell}\right),
    \\
    \epsilon_{\text{init}} &=
    3\left[
    p_0
    +\frac{B_0^2}{2}
    \sech^2\!\left(\frac{y-y_1}{\ell}\right)
    +\frac{B_0^2}{2}
    \sech^2\!\left(\frac{y-y_2}{\ell}\right)
    \right],
\end{align}
with \(\rho_{\rm bg}=0.5\), \(\rho_{\rm sh}=2\), and \(p_0=0.05\).
The transport coefficients used for this run are listed as HS in
Table~\ref{tab:simulations}. 
The computational grid spans $x\in[-100,100]$, $y\in[-200,200]$ and we achieve a Lundquist number $S=\frac{v_A L_x}{r_b}\simeq 20\, 000$ above the tearing instability threshold.  To resolve the ensuing thin current sheet, we use four adaptive mesh refinement levels with a base resolution of $128\times 256$ cells, achieving an effective resolution of $1024\times2048$ cells. 

\begin{figure*}
    \centering
    \includegraphics[width=\textwidth]{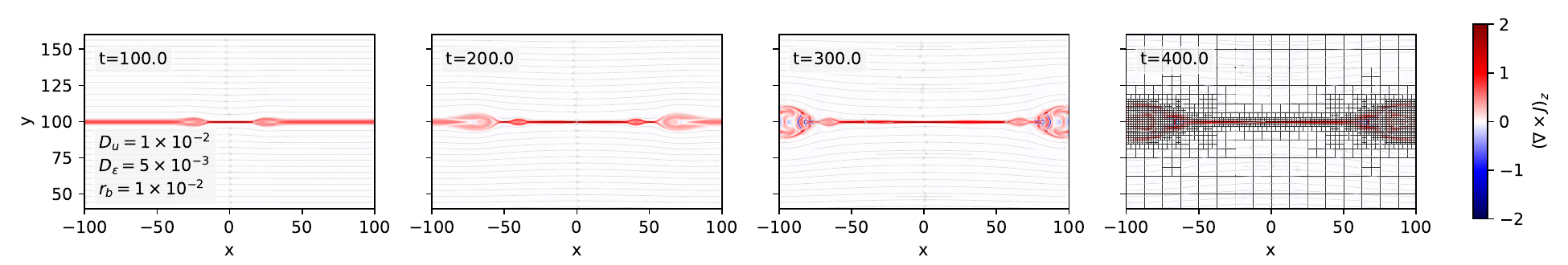}
    \caption{
    Evolution of the out-of-plane current
    \((\nabla\times \mathbf J)_z\) of the top half of the double Harris sheet test at
    \(t=100,200,300,400\). The initial condition consists of two magnetic
    reversal layers located at \(y=L_y/4\) and \(y=3L_y/4\), perturbed by
    the localized vector-potential perturbations in
    \eqref{eq:harris_vector_potential}.    The rightmost panel additionally indicates the adaptive mesh blocks consisting of $16\times16$ cells each.}
    \label{fig:harris_Jz}
\end{figure*}

We first visualize the out-of-plane current
\begin{equation}
    (\nabla \times \mathbf J)_z
    =
    \partial_x J^{ty}
    -
    \partial_y J^{tx}.
\end{equation}
This quantity directly tracks the current sheets. At early times the
current is concentrated around the two magnetic reversal layers. As the
localized perturbation grows, the sheets become distorted and current
concentrates near the reconnecting regions. The evolution shown in
Fig.~\ref{fig:harris_Jz} therefore provides a direct diagnostic of the
formation and subsequent deformation of the current layers.

\begin{figure*}
    \centering
    \includegraphics[width=\textwidth]{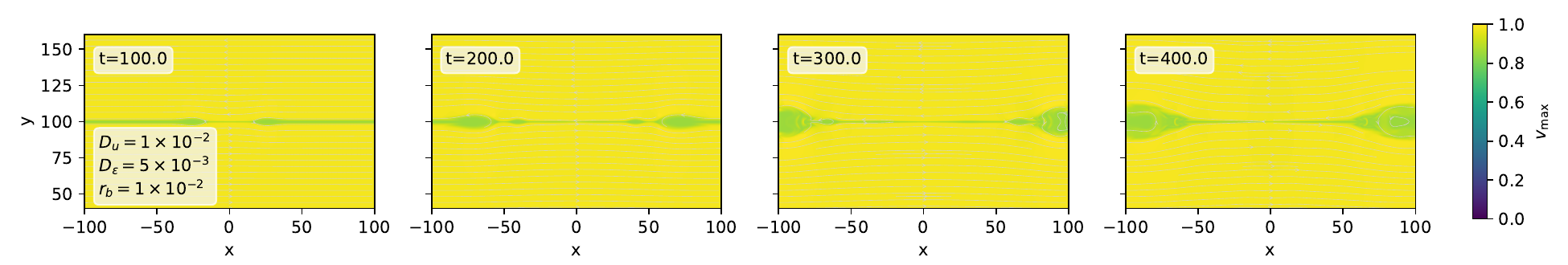}
    \caption{
    Evolution of the maximum local front velocity \(v_{\max}\) for the
    same Harris sheet run as in Fig.~\ref{fig:harris_Jz}. The quantity
    \(v_{\max}\) is computed from the Alfv\'en and magnetosonic front
    velocities described in App.~\ref{sec:front_velocities}, maximizing
    over propagation direction and characteristic branches.
    }
    \label{fig:harris_vmax}
\end{figure*}

We also monitor the maximum local front velocity \(v_{\max}\), which is obtained from the Alfv\'en and
magnetosonic sectors after maximizing over propagation direction and over
the different branches (see App.~\ref{sec:front_velocities}).The result is shown in Fig.~\ref{fig:harris_vmax}. The current sheets and the surrounding
magnetically dominated regions approach the luminal value whereas near the current sheet the maximum front velocity is reduced to $\simeq 0.8$. The global spatio-temporal maximum for the front velocity obtained in the simulation is given by $v_{\text{max}} = 0.9990187$ and the evolution therefore proceeds in a fully causal way.  

\section{Discussion}

We have presented a finite-volume scheme for first-order viscoresistive
relativistic magnetohydrodynamics in which viscous and resistive
corrections are incorporated directly into the constitutive equations.
The resulting model gives a causal extension of dissipative one-form MHD
while retaining the same basic set of hydrodynamic variables. This makes
it possible to evolve energy, momentum, and magnetic flux with
dissipative corrections without introducing independent relaxation
equations for additional nonhydrodynamic fields.
\newline 
In simple limits, the BDNK terms controlled by \(\tau_\epsilon\),
\(\tau_u\), and \(\tau_b\) convert the diffusive equations for energy,
momentum, and magnetic flux into telegrapher-type equations. However,
for the coupled ultra-relativistic energy--momentum sector, this
telegrapher-like structure by itself is not sufficient. The relative sign
structure of the energy and momentum equations implies that, with only
these three BDNK terms, one channel acquires a sign that makes it incompatible with a telegrapher-type structure that allows for stable and causal evolution. The coefficient
\(\tau_X\) resolves this obstruction and is therefore essential for
obtaining a stable and causal evolution of the coupled system, as confirmed by our numerical experiments.
\newline 
The numerical tests demonstrate the robustness of the scheme across a
range of dissipative regimes. The boosted telegrapher benchmark verifies
the implementation of the magnetic diffusion sector and shows the
expected convergence. The shock tube tests show that the method remains
stable in the presence of sharp gradients and discontinuous initial data for a wide range of dissipative parameters, ranging from close to ideal to diffusion-dominated.
The Kelvin--Helmholtz and Orszag--Tang tests further illustrate the
separate roles of viscosity and resistivity in two-dimensional dynamics:
momentum diffusion suppresses shear-driven instabilities, while
resistivity relaxes the coupling between the fluid and magnetic field
lines.
\newline 
The efficiency of the implementation is closely tied to the structure of
\eqref{eq:primitive_recovery}. The BDNK corrections make the conserved
variables depend linearly on the primitive-variable time derivatives, so
that the primitive evolution can be obtained by a local matrix inversion whose entries can be extracted by running the subroutines for the constitutive equations with null and unit vectors. 

While this work focuses on the ultra-relativistic regime (negligible rest-mass contribution), there is no fundamental hindrance to extending the scheme to include particle rest-mass contributions or conserved currents via continuity equation(s) as in \cite{lier2025resistiverelativisticmagnetohydrodynamicsamperes}.  At the same time, the ultra-relativistic dissipative fluid equations discussed here can serve as valid approximations in ultrahigh-energy-density environments such as weakly coupled quark-gluon plasma \citep[][]{RomatschkeRomatschke2007,HeinzSchenke2024,BleicherBratkovskaya2025}, the radiation dominated era in the early universe \citep[][]{Baumann2022} and magnetar magnetospheres with field-strength above the critical Schwinger field strength of $B_*=\frac{m_e^2c^3}{e\hbar}\simeq 4\times 10^{13}\rm G$  \citep[][]{Uzdensky2011}.  In the latter case, dissipation of the magnetic field via reconnection is expected to self-generate a radiation dominated pair plasma, rendering the dynamics both ultra-relativistic and highly collisional.  An interesting application of the presented scheme will hence be to study the dynamics of magnetar flares which involve reconnecting current sheets similar to the ``Harris sheet'' setup discussed in Sec. \ref{sec:harris} \citep[][]{MahlmannPhilippovEtAl2023,ChatterjeePhilippovEtAl2026}.  Further applications should capitalize on the ability to explicitly control both resistive and viscous scales independently -- thereby fixing the magnetic Prandtl number -- which is crucial for correctly capturing complex phenomena such as magnetic dynamos \citep[e.g.][]{BrandenburgSubramanian2005}.

A natural next development is to extend the numerical scheme to include general relativistic effects. \textcolor{black}{In this context, implementations of BDNK are scarce: BDNK has been evolved on $S^2$ in genuine $2+1$-dimensional surface flows using linear-mode, Gaussian-pulse and Kelvin--Helmholtz tests \cite{Keeble:2025bkc}, while compact-star applications have mostly reduced the dynamics to radial ordinary or partial differential problems, including nonlinear $1+1$ spherical neutron-star evolutions in the Cowling approximation \cite{Shum:2025jnl}, linear radial oscillations and collapse-threshold studies \cite{Keeble:2026bzo}, axial quasinormal-mode calculations and viscosity-driven mode families \cite{Bussieres:2026rnz}, and gravitational-wave scattering/absorption or superradiant-amplification calculations for nonrotating and slowly rotating viscous stars \cite{Boyanov:2024jge,Redondo-Yuste:2025ktt}. In the context of MHD, extension to curved space-times is especially relevant for black hole accretion, where strong
gravitational fields, relativistic velocities, magnetic turbulence, and
dissipative transport all enter simultaneously \cite{Del_Zanna_2007,PorthOlivares2017}, as well as for neutron stars in strong magnetic environments. The primitive-variable recovery
structure of \eqref{eq:primitive_recovery} is promising in this
respect, since it provides an economical route for incorporating causal
viscoresistive corrections into general-relativistic MHD simulations. It would be interesting to study the effect of magnetic fields in the strong gravity settings discussed above.}

\begin{acknowledgments}
\noindent
\textit{Acknowledgments---}%
We would like to thank Lorenzo Gavassino, Akash Jain, Pavel Kovtun, Elias Most and Dmitri Uzdensky for helpful discussions. JA is partly funded by the Dutch Institute for Emergent Phenomena (DIEP) cluster at the University of Amsterdam via the DIEP programme Foundations and Applications of Emergence (FAEME). This work was funded by the NWA ORC programme Emergence at All Scales.
\end{acknowledgments}

\appendix
\section{Connecting coefficients to general magnetohydrodynamic theory}
\label{eq:connecting} 
Comparing the contributions in \eqref{eq:firstorderequations} and \eqref{eq:BDNKterms} to the general magnetohydrodynamic theory formulated in Ref.~\cite{Armas_2022}, we find
\begin{equation}
 \begin{split}
&\varepsilon_1=-\tau_\epsilon w~,~\varepsilon_2=-\tau_\epsilon Ts~,~\varepsilon_3=-\tau_e c_T~,~\varepsilon_4=-\tau_\epsilon c_\mu\,\\
 &f_1=\zeta+\frac{\eta}{3}-\tau_X w~,~f_2=\zeta-\frac{2\eta}{3}-\tau_X Ts~,\\
 &f_3=-\tau_X c_T~,~f_4=-\tau_X c_\mu~,\tau_1=\zeta-\frac{2\eta}{3}-\tau_X w~,\\
 &\tau_2=\zeta+\frac{4\eta}{3}-\tau_X Ts~,~\tau_3=-\tau_X c_T~,~\tau_4=-\tau_X c_\mu~,\\
 &\chi_1=\sigma-\tau_u w~,~\chi_2=\frac{\tau_u \mu}{T}~,~\ell_1=\eta~,~k_1=-\tau_u\mu~,\\
 &k_2=\sigma-\tau_u w~,~\rho_1=-\tau_b \mu~,~\rho_3=-\frac{\tau_b \mu}{T}~,\\
 &\rho_4=-\tau_b T~,~m_1=r_\perp~,~n_1=-\tau_b\mu~,~n_2=-\tau_b~,\\
 &\eta_\perp=\eta~,~r_{||}=r_{||}~,
 \end{split}   
 \end{equation}
  while the remaining ones in \cite{Armas_2022} vanish and we defined $\mu=\sqrt{b^2}$, $c_T=d\epsilon/dT +\mu^2/T$ and $c_\mu=T\mu$~.
\section{Front velocities}
\label{sec:front_velocities}
In this Appendix we describe the front velocities used to monitor
causality for nonzero magnetic field. We define
\begin{equation}
    x = W^2 ,
\end{equation}
where \(W\) is the front velocity. Each root \(x\) therefore gives the two
front velocities
\begin{equation}
    W = \pm \sqrt{x}.
\end{equation}
We decompose the magnetic field into components parallel and perpendicular
to the propagation direction,
\begin{equation}
    b_\parallel = b\cos\theta ,
    \qquad
    b_\perp = b\sin\theta .
\end{equation}
For the ultrarelativistic equation of state we introduce the enthalpy $  w=\epsilon+p=\frac{4}{3}\epsilon $ and we define
\begin{equation}
    \mathcal W = w+b^2 .
\end{equation}
The Alfv\'en sector is determined by the two roots
\begin{align}
    \begin{split}
 & x_{\rm A}^{(\pm)}
=
\frac{
\alpha_{\rm A} r_{\rm A}
+ w D_u \tau_b
+ \tau_u \tau_b b_\parallel^2}{
2\tau_b\alpha_{\rm A}
}
 \\  & \pm
\frac{\sqrt{
\left(
\alpha_{\rm A} r_{\rm A}
+ w D_u \tau_b
+ \tau_u \tau_b b_\parallel^2
\right)^2
-4\tau_b\alpha_{\rm A}wD_u r_{\rm A}
} }{
2\tau_b\alpha_{\rm A}
},
  \end{split}
\end{align}
where
\begin{equation}
    \alpha_{\rm A}
    =
    \mathcal W\tau_u-wD_\epsilon ,
    \qquad
    r_{\rm A}
    =
    r_b\,\frac{w+b_\perp^2}{\mathcal W}, 
\end{equation}
and we already imposed  $\tau_{\epsilon} := 2 \tau_{u} $. The corresponding Alfv\'en front velocities are
\begin{equation}
    W_{\rm A}
    =
    \pm \sqrt{x_{\rm A}^{(+)}},
    \qquad
    \pm \sqrt{x_{\rm A}^{(-)}} .
\end{equation}

The magnetosonic sector is determined by a quartic equation in \(x\),
which we write as
\begin{equation}
    \det \mathcal M_{\rm ms}(x)=0 .
\end{equation}
The matrix entering this determinant is
\begin{widetext}
\begin{align}
\begin{split}
 &   \mathcal M_{\rm ms}(x)
=   \\
&
\begin{pmatrix}
\dfrac{\tau_u-D_\epsilon}{3}
+2\tau_u x
&
3\tau_u b_\parallel b_\perp \sqrt{x}
&
wD_\epsilon
-3\tau_u(w+b_\perp^2)\sqrt{x}
&
\tau_u b_\perp(1+2x)
\\[1.0em]
0
&
-wD_u
+\left[\tau_u(w+b_\parallel^2)-wD_\epsilon\right]x
&
-\tau_u b_\parallel b_\perp x
&
\tau_u b_\parallel \sqrt{x}
\\[1.0em]
\dfrac{D_\epsilon-\tau_u-3\tau_X}{3}\sqrt{x}
&
-b_\parallel b_\perp\tau_X
-\tau_u b_\parallel b_\perp x
&
\tau_X(w+b_\perp^2)-2wD_u
+\left[\tau_u(w+b_\perp^2)-wD_\epsilon\right]x
&
-b_\perp(\tau_u+\tau_X)\sqrt{x}
\\[1.0em]
\dfrac{r_b b_\perp}{3\mathcal W}
&
\tau_b b_\parallel \sqrt{x}
&
-\tau_b b_\perp \sqrt{x}
&
-\dfrac{w r_b}{\mathcal W}
+\tau_b x
\end{pmatrix}.
\end{split}
\end{align}
\end{widetext}
Equivalently,
\begin{equation}
    \det \mathcal M_{\rm ms}(x)
    =
    a_4x^4+a_3x^3+a_2x^2+a_1x+a_0 .
\end{equation}
The four magnetosonic roots are denoted by
\begin{equation}
    x_{\rm ms}^{(i)},\qquad i=1,\ldots,4,
\end{equation}
and the corresponding front velocities are
\begin{equation}
    W_{\rm ms}
    =
    \pm\sqrt{x_{\rm ms}^{(i)}} ,
    \qquad i=1,\ldots,4 .
\end{equation}
The maximum front velocity $ v_{\max}$ is then defined as
\begin{equation}
    v_{\max}
    =
    \max_{\theta,\;{\rm branches}}
    \left|
    \operatorname{Re} W(\theta)
    \right| .
\end{equation}

\end{document}